\documentclass[12pt,a4paper,dvips,rotate]{article}
\usepackage{epsfig,wrapfig,times,mathptm} 
\renewcommand{\section}[1]{\vspace{6pt} \noindent\mbox{#1} \newline \noindent}
\renewcommand{\subsection}[1]{\vspace{6pt} \noindent\mbox{\underline{#1}} 
\newline \noindent}
\renewcommand{\subsubsection}[1]{\vspace{6pt} \noindent\mbox{\underline{#1}}
\noindent}

\newfont{\sansb}{cmssbx10}
\newfont{\sans}{cmss10}

\begin{document}

{\center \ {\large THE CAT IMAGING TELESCOPE}
\vspace{30pt}\\

A.~Barrau$^5$, R.~Bazer-Bachi$^2$, H.~Cabot$^7$, L.M.~Chounet$^4$, G.~Debiais$^7$,
B.~Degrange$^4$, J.P.~Dezalay$^2$, A.~Djannati-Ata\"{\i}$^5$, 
D.~Dumora$^1$, P.~Espigat$^3$, B.~Fabre$^7$, P.~Fleury$^4$, G.~Fontaine$^4$, R.~George$^5$,
C.~Ghesqui\`ere$^3$, P.~Goret$^6$, C.~Gouiffes$^6$, I.A.~Grenier$^{6,9}$, L.~Iacoucci$^4$,
S.~Le Bohec$^3$, I.~Malet$^2$, C.~Meynadier$^7$, F.~Munz$^8$, T.A.~Palfrey$^{10}$, E.~Par\'e$^4$, 
Y.~Pons$^5$, M.~Punch$^3$, J.~Qu\'ebert$^1$, K.~Ragan$^1$, C.~Renault$^{6,9}$, 
M.~Rivoal$^5$, L.~Rob$^8$, P.~Schovanek$^{11}$, D.~Smith$^1$,  J.P.~Tavernet$^5$
and J.~Vrana$^{4 \dag}$ \vspace{6pt}\\
{\it $^1$Centre d'Etudes Nucl\'eaire de Bordeaux-Gradignan$^{\ast}$, France \\
$^2$Centre d'Etudes Spatiales des Rayonnements$^{\ddagger}$, Toulouse, France \\ 
$^3$Laboratoire de Physique Corpusculaire$^{\ast}$, Coll\`ege de France, Paris, France \\
$^4$Laboratoire de Physique Nucl\'eaire de Haute Energie$^{\ast}$, Ecole Polytechnique, Palaiseau, France \\
$^5$Laboratoire de Physique Nucl\'eaire de Haute Energie$^{\ast}$, Universit\'es de Paris VI et VII, France \\
$^6$Service d'Astrophysique$^{\#}$, Centre d'Etudes de Saclay, France \\
$^7$Groupe de Physique Fondamentale$^{\ast}$, Universit\'e de Perpignan, France \\
$^8$Nuclear Center, Charles University, Prague, Czech Republic \\
$^9$Universit\'e Paris VII \\
$^{10}$Department of Physics, Purdue University, Lafayette, IN 47907,
U.S.A \\
$^{11}$JLO Ac. Sci. \& Palacky University, Olomouc, Czech Republic \\
$^{\ast}$IN2P3/CNRS \\
$^{\ddagger}$INSU/CNRS \\
$^{\#}$DAPNIA/CEA \\
$^{\dag}$Deceased \\
}}

 {\vspace{12pt}}
{\center ABSTRACT\\}
The VHE gamma-ray imaging telescope CAT started taking data in October 1996. 
Located at the Themis solar site in 
southern France (2$^\circ$E, 42$^\circ$N, 1650 m a.s.l), 
it features a 17.7 m$^2$ Davies-Cotton mirror equipped  
with a 600 PMT camera at the focal plane. The mechanics and optics,
the PMTs and the
electronics are presented. The performance, based on the first 7 months of
operation, is described.                                                       

\setlength{\parindent}{1cm}
\section{INTRODUCTION}
A new atmospheric Cherenkov imaging telescope (Figure 1), designed for VHE 
gamma-ray astronomy,
has been operating since October 1996. It was built by a Franco-Czech 
collaboration and installed at the same site, at Themis, 
as the two existing  timing arrays 
ASGAT (Goret et al., 1993)  and THEMISTOCLE (Djannati-Ata\"{\i} et al., 1995). 
The ensemble of
these three instruments, using complementary detection techniques, has been named
CAT for Cherenkov Array at Themis. The heart of the imaging telescope 
consists of a fine grained camera with a pixel size of the order of the width
of gamma-ray images. The different parts of the telescope are described  below. 

\section{THE REFLECTOR}
The Davies-Cotton reflector of the CAT imager consists of 90 spherical mirrors
50 cm in diameter with a radius of curvature of 12$^{+0.24}_{-0.0}$ m. 
The total collecting area is 17.7m$^2$ and the focal length is 6 m. 
The weight and rigidity
of the supporting structure were optimized by using state of the
art design and manufacturing.
The 1 cm thick individual glass mirrors
are first surface aluminized and protected by SiO$_{2}$. They are positioned on the
frame according to their measured radii of curvature and
aligned to better than 0.1 mrad using an autocollimation technique. 
Regular checks of mirror alignment are
performed. The focal spot
obtained, when imaging stars onto a white sheet, is 
of the order of 1.8 mrad fwhm. The relative time delay across the mirror 
resulting from the Davies-Cotton design is 1.6 ns from center
to edge.
The encoding system of the 
alt-azimuth mount of the telescope allows a pointing resolution of 
$\sim$0.14 mrad in both azimuth and elevation. The pointing corrections, needed
for an accurate determination of the source position in the focal plane, 
are described later in this paper.

\section{THE FOCAL PLANE DETECTOR}
The layout of the focal plane detector is shown in Figure 2. An inner zone consists
of a densely packed array of PMTs with a center to center spacing of
13 mm equivalent to 2.2 mrad. This zone, used for the fine imaging of Cherenkov 
showers, is filled with 546 fast 11 mm diameter PMTs (Hamamatsu R1635-02) .
The full  field of view (f.o.v) of the inner zone is 3.1$^\circ$ in diameter.
An outer double ring of 54 larger PMTs, 28mm diameter (Hamamatsu R6076)  
extends the  f.o.v. to 4.8$^\circ$ with a coarse granularity.
All 600 PMTs are equipped
with aluminized Winston cones (Punch, 1994)
in order to both reduce the dead space between
photocathodes and cut off unwanted stray light. Measurements of the
transmission of light to the photocathode confirm the sharp cutoff for
incident angles on the Winston cones greater than $\sim$32$^\circ$.
With the use of such cones,
the light collection efficiency is nearly doubled
relative to the no-cone situation.

\section{PMTS AND ELECTRONICS}
The trigger electronics of the CAT telescope has been designed to be placed immediately
behind the PMTs, allowing it to take advantage of the brevity of the
Cherenkov pulse, the isochronism of the mirror, and the speed of the
PMT response. 
A detailed description of the electronics can be found in Barrau (1997).

\subsection{Photomultiplier Tubes}
The Hamamatsu R1635 PMTs typically show 
a risetime of $\sim$0.9 ns and a width of $\sim$1.5 ns.
A laboratory calibration was performed which yielded, first 
the gain as a function of high voltage ,then
the standard deviation $\sigma$ and the mean value Q of the
single photoelectron (p.e.) peak for each PMT. The value of $\sigma$/Q is 
0.45$\pm$0.05 for a peak/valley ratio of $\sim$2, allowing for a straightforward
measurement of the gains.
In the experiment, the PMTs are operated at a gain of 10$^6$ at a 
high voltage between 1050V and 1200V.
The high voltage is supplied by a 
computer controlled CAEN SY527 generator. 
The gains are measured every month, using a pulsed LED at very low light
level, 
and observing the single p.e. peak.
After shaping by an
OPA623 amplifier, PMT signals are split into two channels, one for the
trigger generation and scaler updating and the other for charge measurement.
The trigger and scaler channels are further amplified ($\times$15) with
NEC1678 wide-band amplifiers.

\subsection{Trigger Generation and Scalers}
The PMT signals are discriminated
using comparators. The thresholds are adjusted by software with a minimum value
of $\sim$1 photoelectron (at a gain of 10$^6$).
The inner 288 PMTs participate in the trigger
generation (see Figure 2). The large combinatorial factor is avoided
by dividing the trigger zone 
into 
9 angular sectors of 48 PMTs, with an overlap of
16 PMTs between adjacent sectors, to prevent a loss of trigger efficiency 
at the boundaries. A majority logic trigger is formed, within each
sector, requiring p pixels out of 48 above n photoelectrons. A final 'OR' of the
outputs of all
sectors provides the camera trigger. Comparator outputs are also sent to 100 MHz
scalers to monitor the PMT count rates.

\subsection{Charge Measurement}
After a 140 ns delay cable, the charge signals are gated by 12 ns -wide fast analog
switches opened by the camera trigger. Thus the contribution of the night sky
background light to the charge measurement is minimized. The signals are
then amplified ($\times$4) and analyzed by 15 bit Fastbus ADC's (Lecroy 1885).
These ADC's feature a resolution of 50 fC/channel for signals up to 200 pC and
of 400 fC/channel above. The effective dynamic range runs from $\sim$2
p.e./pixel up to $\sim$1000 p.e./pixel and the conversion factor is typically 12 
counts/p.e..

\section{POINTING CORRECTIONS}
In order to fully exploit the high definition of the CAT imaging camera, the
position of the observed source in the camera has to be known to better than
0.5 mrad ($\sim$1/4 of the pixel size).
However, misalignment of the rotation
axes, mechanical deformations and other effects lead to pointing errors
amounting to a few mrads. 
The true pointing position is measured by means of two CCD
cameras. A first one, located at the center of the mirror, views the focal
plane. Three green LED's, one at the center and the others at the
outer edge of the camera, provide an absolute reference relative to the camera
coordinates. A second CCD camera, located near the first one and
coaligned with the mirror axis, views the
same sky region as the PMT camera. 
The information from both CCDs enables to localize the position of the
observed source in the PMT camera with an accuracy of $\sim$0.3
mrad. An
independant cross-check was performed by analyzing the 
track of bright stars through the PMTs 
with the telescope pointing at fixed positions in azimuth
and elevation. The pointing accuracy achieved with this method is $\pm$0.2
mrad.
A third analysis involved the identification of the star field in 
the PMT camera during  observations.
All three methods give consistent results leading to
off-line pointing corrections accurate to better than 0.3 mrad.

\section{PERFORMANCE}
As of the writing of this paper, only the inner 546 small PMTs have been mounted
in the camera. After 7 months of operation, 10 of them are not working 
properly. 
A small drift in PMT gains was detected and corrected for. All
scientific data have been recorded with a trigger level of $\ge$ 4 pixels passing a
threshold of
3 photoelectrons. The corresponding trigger rate at zenith is of the
order of 15-20 Hz with a negligible random coincidence rate. The muon trigger
rate is $\sim$6 Hz as was measured under cloudy conditions. 
The night sky background light is routinely monitored using the 
shape of the ADC
pedestals for random gates. It is fairly constant from night to night at a level of
$\sim$0.015-0.020 p.e./ns/pixel. The singles rates
are of the order of a few kHz  except for those PMTs
which see a star. Whenever the singles rate for a given pixel reaches
$\sim$7Mhz, as is the case when a bright star enters a PMT, the HV for this
pixel is automatically lowered  to avoid excessive anode current.
The effective
energy threshold for gamma rays is estimated at $\sim$220 GeV at 20$^\circ$
from zenith.
Joint observations of Mrk501 with the THEMISTOCLE experiment are
under analysis to verify the absolute energy calibration.
\noindent
The good sensitivity of the CAT imager is demonstrated by the detection 
of the Crab
nebula and of the blazars Mrk421 and Mrk501.

\section{CONCLUSIONS}
The VHE gamma-ray imaging telescope CAT has been taking data since October 1996.
Its 546 pixel camera makes it the finest grained imager operating to-date. The
mechanical structure and optics have met specifications to give a 
focal spot matching the
2.2 mrad pixel size. The dedicated electronics, tailored to the
use of both fast and numerous PMTs, have met the requirements of good trigger
efficiency, accurate charge measurement and reliability. Winston cones have proven
to be efficient for eliminating stray light so that single rates are quite
stable. Finally the overall sensitivity of the equipment has been checked by
the positive detection of the Crab nebula ,  Mrk421 and Mrk501.
Future improvements concern the implementation of a guard ring of 54 PMTs for
more accurate energy determination and of a muon barrel for studying the
response of the telescope to muons.

\section{REFERENCES}
\setlength{\parindent}{-5mm}
\begin{list}{}{\topsep 0pt \partopsep 0pt \itemsep 0pt \leftmargin 5mm
\parsep 0pt \itemindent -5mm}
\vspace{-15pt}
\item Barrau, A., Nucl. Inst. and Meth. in Phys. Res. A, 387, 69 (1997).
\item Djannati-Ata\"{\i}, A. Proc. 24th ICRC, 2, 315 (1995).
\item Goret, P., Palfrey, T., Tabary, A. et al., A\&A, 270, 401 (1993).
\item Punch, M., in Towards a Major Atmospheric Cerenkov Detector III, ed. T.
Kifune, pp. 215-220, Universal Academy Press Inc., Tokyo (1994)
\end{list}

\begin{figure}[hb]
 \begin{center}
   \mbox{\epsfig{file=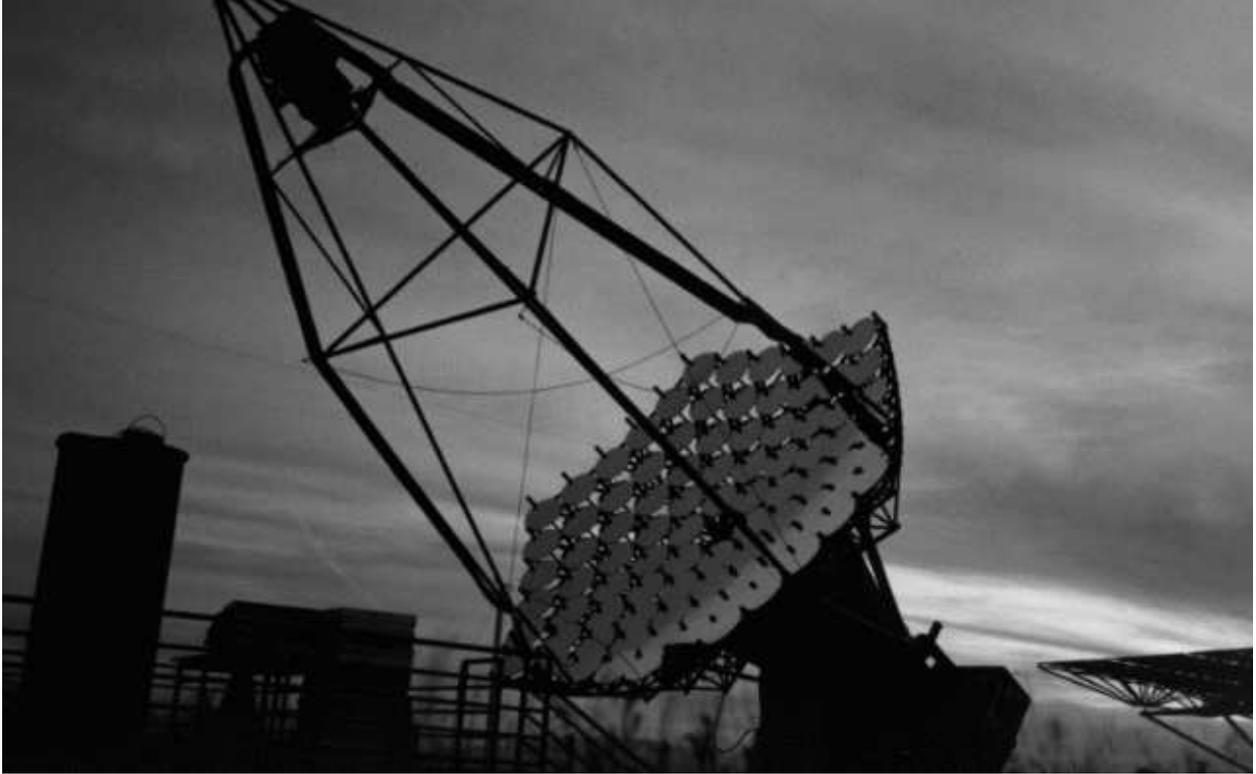,height=10.5cm}}
 \end{center}
 \caption{The CAT imaging telescope}
 
\end{figure}

\begin{figure}[ht]
 \begin{center}
   \mbox{\epsfig{file=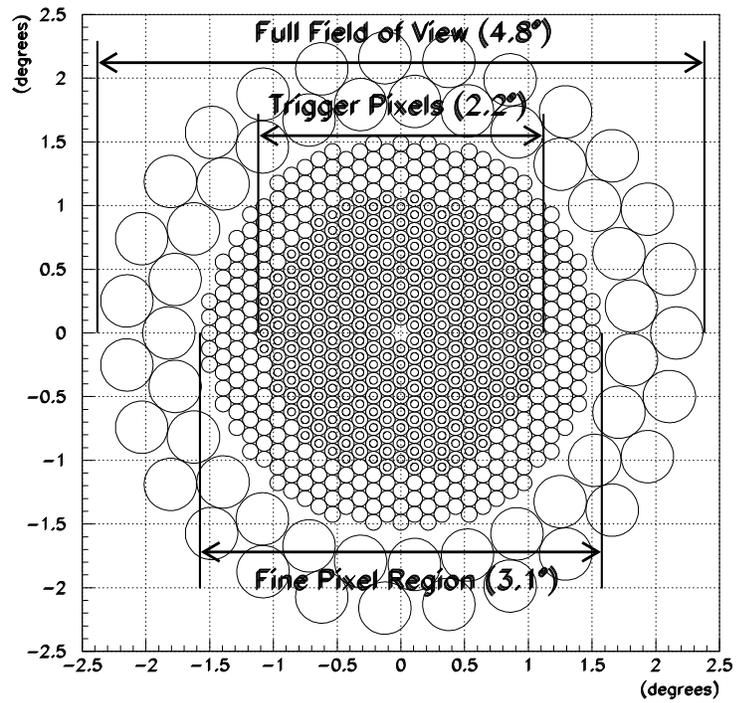,height=10.5cm}}
 \end{center}
 \caption{The very-high-definition camera. The spacing in the fine-pixel region
 is 0.12$^{\circ}$. The 288 inner PMs, shown filled, are used in the trigger
 logic.}
\end{figure}

\end{document}